\documentclass[a4paper,11pt]{article}
                                                                                                                                                                                                                                                                                                                                                                                                                                                                                                                                                           
\usepackage{amssymb,amsthm,amscd, amsbsy, array,amsmath,mathdots}
\usepackage{cancel}
\usepackage[francais]{babel}
\usepackage{fontenc}
\usepackage{amsfonts}
\usepackage[mathscr]{eucal}
\usepackage{mathrsfs}
\usepackage{latexsym}
\usepackage{graphicx}
\usepackage[a4paper,textwidth=17cm,textheight=21cm]{geometry}

\newcommand{\be}{\begin{equation}}
\newcommand{\ee}{\end{equation}}
\newcommand{\bea}{\begin{eqnarray}}
\newcommand{\eea}{\end{eqnarray}}

\def\1{{\mbox{\boldmath $1$}}} %
\newcommand{\Z}{\mathbb{Z}}
\newcommand{\R}{\mathbb{R}}
\newcommand{\C}{\mathbb{C}}

\newtheorem{proposition}{Proposition}

\newtheorem{theorem}{Theorem}
\newtheorem{defin}{Definition}

\begin{document}
\thispagestyle{empty}
\begin{center}
	\begin{huge}
		Noncommutative geometry of the dihedral group $D_6$
	\end{huge}
\end{center}

\begin{center}
	\begin{large}
		ARM Boris Nima \\
	\end{large}
	{\footnotesize  \textit{arm.boris@gmail.com}}
\end{center}

\vspace{5cm}
\begin{center}
	\textbf{Abstract} \\
We study the noncommutative geometry of the dihedral group $D_6$ \\
using the tools of quantum group theory. We explicit the \\
torsion free regular spin connection and  the \\
 corresponding 'Levi-Civita' connection. \\
 Next, we find the Riemann curvature \\
 and its Ricci tensor. The main \\
  result is the Dirac operator \\
  of a representation of the \\
  group which we find the \\
   eigenvalues and the \\
   eigenmodes. 
\end{center}

\newpage

\setlength{\textheight}{21cm}
\addtolength{\voffset}{0.2cm}

\section{Introduction}
We study the noncommutative Riemannian geometry of the Dihedral group $D_6$ which are the rotation and the symmetries leaving invariant the haxagon.
We used the quantum group method to study this noncommuative Riemannian geometry.
\paragraph{}
First we introduce the mathematical tools of quantum groups. We start with the right translation operator which is related to the derivation on the group. We explicit the Maurer-Cartan  equation  and covariant derivative of 1-forms. We continue with 
the relation between weighted product of spin connection. Then we define the Riemann curvature and the Ricci tensor associated to quantum groups methods.  After defining the lift, we define the Dirac operator depending on the spin connection. 
\paragraph{}
We choose a cyclic class conjugaison defined in Definition 1 of the Dihedral group $D_6$.
With the conjugaison action table (Table 1) of $D_6$, we can deduce the relation between 2-forms. Next, we choose a basis of the space $\Omega^2(H)$ spanned by 2-forms.
With relation between spin connection and two forms, we can find the torsion free connection (\ref{definitiondeA}). With the multiplication table (Table 2), we can find which relations obey the regular spin connection. Furthermore, we calculate Ad-invariant bilinear form (which the identity) and its associated metric. Moreover, we calculate the relation of right translation of elements of the conjugacy class function of the parameter of the regular spin connection.
\paragraph{}
By the way, we fully explicit the elements of the Dihedral group $D_6$ and its cyclic conjugacy class. In addition, we can totally know the regular spin connection function of the element of the basis of 1-forms. As a result, we find the 'Levi-Civita'  connection and the Riemann curvature for this group. Hence we get the canonical lift and the Ricci tensor of the theory.
\paragraph{}
In fact, we write the representation of the Dihedral group $D_6$. From this, we deduce its Casimir operator and the corresponding gamma matrix. As a  consequence, we find the famous Dirac operator for the group. Because this Dirac operator depend on right translation operator, we can totally, with the help of mathematica,  calcuate the eigenvalues of the Dirac operator. At this point, we want find the eigenmodes of the Dirac operator. To this end, we find the eigenmodes of the corresponding waves operator. Thereby, we use the Peter-Weyl decomposition to find all the eigenmodes of the Dirac operator.
\paragraph{}
 Finally, because we want to apply the spectral action theory, we calculate it for the Dirac operator we have just found.
 \paragraph{}
 To conclude with, we want to say that this paper is the copie of \cite{A4}, which is done for the Alternating group $A_4$, done for the Dihedral group $D_6$.

\newpage
\section{Preliminaries}
\paragraph{}
In this section, we briefly recall the basic definitions of differential structures 
\cite{Woronwicz} and of noncommutative geometry \cite{riemanngeometry}, specialised to the case of finite groups that we need in the paper. Thus, we work with the Hopf algebra $H=\mathbb{C}[G]$ of functions on a finite group $G$. We equip it with its standard basis $(\delta_g)_{g \in G}$ defined by Kronecker delta-functions $\delta_g (h) =\delta_g^h ,\forall g,h \in G$. Let $\mathcal{C}$ be a nontrivial conjugacy class of $G$, and $\Omega_0$ the vector subspace
\begin{equation}
\Omega_0=\{ \delta_a | a\in \mathcal{C}\} =\C \mathcal{C}
\end{equation}
Any  Ad-stable set not containing the identity can be used here, but we focus on the irreductible case of a single conjugacy class. From the Woronowicz's theory in \cite{Woronwicz}, the first order differential calculus associated to $\mathcal{C}$ is generated over $\C [G]$ by $\Omega_0$ and given by 
\begin{equation}
\mathrm{d}f=\sum_{a\in \mathcal{C}} (\partial^a f)e_a,\partial^a =R_a-\mathrm{id},e_a f =R_a(f)e_a
\label{relationcommutation}
\end{equation}
$\forall a \in \mathcal{C}, f \in H$, where the operator $R_a$ is defined by $R_a(f)(g)=f(ga), \forall g \in G$. The Maurer-Cartan 1-form $e:\Omega_0 \to \Omega^1(H)$ is given by
\begin{equation}
e_a=e(\delta_a)=\sum_{g \in G} \delta_g \mathrm{d}\delta_{ga}
\end{equation}
and higher differential forms are obtained from Woronowicz skew-symmetrization procedure 
\cite{Woronwicz}, using the braiding 
\begin{equation}
\Psi (e_a   \underset{H}{\otimes } e_b ) =e_{aba^{-1}} \underset{H}{\otimes } e_a
\end{equation}
The Maurer-Cartan equatino takes the form
\begin{equation}
\mathrm{d}e_a =\theta \wedge e_a +e_a \wedge \theta ,\hspace{3mm}\theta =\sum_{a \in \mathcal{C}} e_a 
\end{equation}
The element $\theta$ obeys $\theta \wedge \theta =0$ and generates  d ingeneral as graded-commutator with $\theta$. Lemma 5.3 in \cite{riemanngeometry} gives the full set of relations of $\Omega^2 (H)$, namely
\begin{equation}
\sum_{a,b \in \mathcal{C}; ab=g} \lambda_a^{g,\beta} e_a \wedge e_b =0, \forall g \in G, \forall \beta
\label{relationaveclambda}
\end{equation}
where for $g \in G$ fixed, $\{\lambda^\beta \}$ is a basis of the invariant subspace of the vector  space spanned by $\mathcal{C} \cap g \mathcal{C}^{-1}$ under the automorphism $\sigma(a)=a^{-1}g$. There are also cubic and higher degree relations (which   are in fact nontrivial in our case of $D_6$) but we will not need them explicitly (most of Riemannian geometry needs only 1-forms and 2-forms).
\paragraph{}
Next, following \cite{riemanngeometry}, a framing means a basis of $\Omega^1(H)$ over $\C[G]$, and an action  of the frame group. In our case  we chose the framing to be components $\{e_a\}$ of the Maurer-Cartan form as above and for frame group we choose $G$ itself, acting by Ad. This is a canonical choice and its classical meaning is explained in 
\cite{quantumbraided}.  Aspin connection is then a collection $\{A_a\}_{a\in \mathcal{C}} \}$ of component 1-forms. Its associated covariant derivative is defined on an  1-form $\alpha=\alphaâ e_a$ by 
\begin{equation}
\nabla \alpha =\mathrm{d}\alpha^a \underset{H}{\otimes } e_a - \alphaâ \sum_{b \in \mathcal{C}} A_b \underset{H}{\otimes } (e_{b^{-1}ab} -e_a)
\label{nablaalpha}
\end{equation}
with summation on $a$. The asociated torsion tensor $T:\Omega^1 (H) \longrightarrow \Omega^2 (H)$ is defined by $T\alpha =\mathrm{d} \wedge \alpha -\nabla \alpha$ and the zero-torsion condition is vanishing 
\begin{equation}
\overline{D}_A e_a \equiv \mathrm{d}e_a +\sum_{b\in \mathcal{C}} A_b \wedge (e_{b^{-1}ab} -e_a)
\label{dbarre}
\end{equation}
The spin connection here has values in the dual space $\Omega^*_0$, which is a 'braided-Lie algebra' in a precise sense. Associated to this geometrical point of view, there is a regularity condition 
\begin{equation}
\sum_{a,b \in \mathcal{C}; ab=g} A_a  \wedge A_b =0, \hspace{3mm} \forall g \neq e, g \notin \mathcal{C}
\label{regularitycond}
\end{equation}
The curvature $\nabla^2$ associated to a regular connection $A$ is in frame bundles terms a collection of 2-forms $\{F_a \}_{a \in \mathcal{C}}$ defined by 
\begin{equation}
F_a=dA_a +\sum_{c,d \in \mathcal{C},cd=a} A_c \wedge A_d -\sum_{c \in \mathcal{C}} (A_c \wedge A_a +A_a \wedge A_c )
\label{F}
\end{equation}
while  the Riemann curvature  $\mathcal{R}: \Omega^1(H) \longrightarrow \Omega^2(H)  \underset{H}{\otimes }  \Omega^1 (H)$ is given by
\begin{equation}
\mathcal{R}\alpha =\alpha^a \sum_{b \in \mathcal{C}} F_b \underset{H}{\otimes }(e_{b^{-1}ab}-e_a)
\label{R}
 \end{equation}
Finally, the Ricci tensor is given by
\begin{equation}
Ricci =\sum_{a,b,c \in \mathcal{C}} i(F_c)^{ab} e_b  \underset{H}{\otimes } (e_{c^{-1}ac}-e_a)
\end{equation}
where $i(F_c)=i(F_c)^{ab} e_a \underset{H}{\otimes } e_b$ and $i: \Omega^2(H) \longrightarrow \Omega^1(H) \underset{H}{\otimes } \Omega^1 (H)$
is
 a lifting which  splits the projection of 
$\wedge$. A canonical choice is \cite{riemanngeometry} 
\begin{equation}
i(e_a \wedge e_b)=e_a \underset{H}{\otimes } e_b -\sum_{\beta} \gamma^{\beta,\alpha} \sum_{c,d \in \mathcal{C}, cd=ab} \lambda^b_c e_c \underset{H}{\otimes } e_d
\end{equation}
where $\{ \gamma^\beta \}$ are the dual basis to the  $\{ \lambda^\beta \}$ with respect to the dot product as vector in $\C \cap ab  \mathcal{C}^{-1}$. Another canonical 'lift' is $i'=\mathrm{id}-\Psi$ but note that in this case $i' \circ \wedge$ is \textit{not}
a projection operator.
There are two further structures that one may impose in this situation. First of all, given  a choice of framing $\{e_a \}$, i.e. again a basis of $\Omega^1$ but now as a right 
$\C[G]$-module, and transforming under the contragradient action of $G$ (the corresponding 
metric is $g=\sum_a e^{*a} \underset{H}{\otimes } e_a$). The cotorsion of a spin connection is the torsion with respect to the coframing, and is given by 
\begin{equation}
D_A e^{*a} \equiv \mathrm{d}e_a +\sum_{b\in \mathcal{C}} A_b \wedge (e^{*bab^{-1}} -e^{*a})
\label{DAetoile}
\end{equation}
Vanishing of contorsion  has the classical meaning of a generalisation of metric compatibility of the spin connection, see \cite{quantumbraided}. So we are usually interested in regular torsion-free and cotorsion free  connections.
\paragraph{}
Finally, a 'gamma-matrix' is defined \cite{riemanngeometry} as a collection of endomorphism $\{ \gamma_a \}_{a \in \mathcal{C}}$ of a vector space $W$ on which $G$ acts by a representation $\rho_W$, a 'spinor field' is a $W$-valued function on $G$ and the Dirac operator  on the spinor fields is
\begin{equation}
\cancel{D}=\partial^a \gamma_a -A^b_a \gamma_b  \tau^a_W
\label{defdiracoperator}
\end{equation}
where $A_b =A_b^a e_a$ and $\tau^a_W =\rho_W(a^{-1}-e)$. There is a canonical choice where  $\gamma$ is built from $\rho_W$ itself, explained in  \cite{riemanngeometry}

\section{Cyclic Riemannian structures}
In this section, we construct Riemannian geometry on groups  endowed  with conjugacy classes which obey a certain cyclicity condition. For the case when the differential calculus is of degree three, we determine the entire exterior algebra and the moduli space of torsion free connections, and for any degree $n \ge 2$, we give the general form of the invariant metric.
\begin{defin}
Let $\mathcal{C}$ be a conjugacy class with $n$ elements, $n \ge 2$, in a group $G$. We say that $\mathcal{C}$ is 'cyclic' if there exists at least one $t$ in $\mathcal{C}$ such that $\mathrm{Ad}_t$ is a cyclic permutation of $\mathcal{C}-\{t\}$ and the map $a \to \mathrm{Ad_a}(t)$ is a permutation of $\mathcal{C}$.
\end{defin}
For $n=3$ we have the following characterisation of $\Omega^2(H)$
\begin{proposition}
For a cyclic conjugacy class $\mathcal{C}=\{t,x,y\}$ of order $3$ in a finite group $G$, the bimodule $\Omega^2(H)$ of 2-form  is 6-dimensional over $\C[G]$ and is defined  by the following equations
\begin{equation}
e_a \wedge e_a=0 \hspace{3mm} \sum_{a,b \in \mathcal{C}, ab=g} e_a \wedge e_b=0
\label{relation2formes}
\end{equation}
$\forall a \in \mathcal{C}, g \in G$, where $(e_a)_{a \in \mathcal{C}}$ is the basis of Maurer-Cartan 1-forms.
\end{proposition}
\textit{Proof}
\\
We assume the existence of an element $t \in \mathcal{C}$ as in Definition 1. Without loss  of generality, we denote the other elements of $\mathcal{C}$ by $x,y,z$ with the following table for Ad:
\\
\begin{center}
\begin{tabular}{|c|c|c|c|}
\hline
Ad & t &x&y\\
\hline
t&t&y&x \\
\hline
x&y&x&t\\
\hline
y&x&t&y\\
\hline
\end{tabular}
\\
\vspace{3mm}
Table 1
\end{center}
It follows that
\begin{equation}
xt=yx=ty ;\hspace{5mm} yt=xy=tx
\label{relation}
\end{equation}
Using relations (\ref{relation}), we apply the Woronowicz antisymmetrization procedure to obtain the following relations of the form  (\ref{relation2formes}) in $\Omega^2(H)$
\begin{eqnarray}
e_a \wedge e_a=0,\hspace{4mm} \forall a \in \mathcal{C}\\
e_x \wedge e_t +e_y \wedge e_x+e_t  \wedge e_y=0 \\
e_y \wedge e_t+e_x \wedge e_y+e_t\wedge e_x=0
\label{equ2formD6}
\end{eqnarray}
This form (\ref{relation2formes}) holds for any group since the elements $e_a \otimes e_a$ and $\sum_{ab=g} e_a \otimes e_b$ are in the kernel of $id-\Psi$. However, using 
\ref{relationaveclambda} one may see that they are the \textit{only}  relations of $\Omega^2(H)$ which is therefore of dimension $6$ as stated.
\begin{flushright}
$\spadesuit$
\end{flushright}
From now, we choose a  basis of $\Omega^2(H)$ to be
\begin{equation}
\{e_x \wedge e_t, e_y \wedge e_x, e_y \wedge e_t, e_x \wedge e_y \}
\label{basis}
\end{equation}
For convenience, we will sometime use  indexes $1,2,3$ to refer  to  $t,x,y$ respectively.

\paragraph{}
In fact for most geometric purposes we need only the exterior algebra up to degree 2, so we limit ourselves to the general  result about dimensions. In principle one may go on to compute explicit relations in higher  degree and a Hodge $*$  operator as in \cite{Strois}
using the metric below, etc. The results is an important reminder  that the degree 2 relations alone may not be enough for a geometrically reasonable exterior algebra.
 \begin{proposition}
 In the setting  of Proposition 1 above and for the framing defined by the Maurer-Cartan 1-form, the moduli space of torsion free connections is $2|G|$-dimensional and is given by the following components 1-forms:
 \begin{eqnarray}
\nonumber A_t=(1+\alpha)e_t+\gamma e_x+\beta e_y\\
\nonumber A_x=\gamma e_t+(1+\beta)e_x+\alpha e_y \\
A_y=\beta e_t+\alpha e_x +(1+\gamma) e_y
\label{definitiondeA}
\end{eqnarray}
where $\alpha, \beta, \gamma$ are functions on $G$ such that 
\begin{equation}
\alpha +\beta +\gamma=-1
\end{equation}
Thus we have also 
\begin{equation}
\sum_{a\in \mathcal{C}}A_a=0
\label{sommedesAzero}
\end{equation}
 \end{proposition}
\textit{Proof}
\\
We follow the same method as for $S_3$ in \cite{riemanngeometry}. In the framing defined by the Maurer-Cartan  1-form, the torsion free connections obey the following equation 
(see (\ref{dbarre}))
\begin{equation}
\sum_{b \in \mathcal{C}} A_b \wedge (e_{b^{-1}ab}-e_a)+\sum_{b\in \mathcal{C}}(e_b\wedge e_a +e_a \wedge e_b)=0
\label{relationconnection2form}
\end{equation}
$\forall a \in \mathcal{C}$. Using Table 1, we write (\ref{relationconnection2form}) as 
\begin{eqnarray}
\nonumber A_x \wedge (e_y-e_t)+A_y\wedge (e_x-e_t)+(e_x+e_y)\wedge e_t+e_t\wedge(e_x+e_y)=0\\
\nonumber A_t \wedge (e_y-e_x)+A_y\wedge (e_t-e_x)+(e_t+e_y)\wedge e_x+e_x \wedge (e_t+e_y)=0\\
A_t \wedge (e_x-e_y) +A_x\wedge (e_t-e_y)+(e_t+e_x)\wedge e_y+e_y \wedge (e_t+e_x)=0
\label{systequ2formavecA}
\end{eqnarray}
We just have to solve the first two equations since the third one in this systemp can be obtained from the other by simple summation. We set $A_a=A_a^b e_b$ (sum over $b \in \mathcal{C}$) for functions $A_a^b  \in H$ with
\begin{equation}
A_t^t=1+\alpha \hspace{3mm}A_x^x=1+\beta \hspace{3mm}A_y^y=1+\gamma
\end{equation}
We put this into the equations  to be solved  and write them in the basis (\ref{basis}). Using the fact that each coefficient of the basis element has to vanish, we obtain
\begin{eqnarray}
A^x_t=\gamma =A^t_x \hspace{3mm} A^x_y=\alpha =A^y_x\\
A^t_y=\beta =A^y_t\\
\alpha +\beta +\gamma=-1
\end{eqnarray}
as stated. Finally using these solutions one checks by simple computation that $A_t+A_x+A_y=0$.
\begin{flushright}
$\spadesuit$
\end{flushright}
\paragraph{}
We now study the regularity of connections
\begin{proposition}
Under the hypothesis of Proposition 1 the regular connections are either solutions of the system:
\begin{eqnarray}
\nonumber A_t \wedge A_t=0, A_x \wedge A_x=0, A_y \wedge A_y=0\\
\nonumber A_x \wedge A_t+A_y \wedge A_x +A_t \wedge A_y=0\\
A_t \wedge A_t +A_x \wedge A_x +A_y \wedge A_y=0
\label{nonlinearsystem}
\end{eqnarray}
\end{proposition}
\textit{Proof}\\
The general form of the regularity's aquation is given by (\ref{regularitycond}). One then needs the multiplication table at least for the elements of the class $\mathcal{C}$, by enumeration of the cases we find that under the hypothesis of Proposition 1, the only possible case is shown in Table 2. These correspond to the possibility stated.
\begin{flushright}
$\spadesuit$
\end{flushright}
\begin{center}
\begin{tabular}{|c|c|c|c|}
\hline
$\times$  & t &x&y\\
\hline
t&$t^2$&yt&xt \\
\hline
x&xt&$t^2$&yt\\
\hline
y&yt&xt&$t^2$\\
\hline
\end{tabular}
\\
\vspace{3mm}
Table 2
\end{center}
\paragraph{}
To explicitly solve this nonlinear system (\ref{nonlinearsystem}) one needs more precision on the group $G$. We solve (\ref{nonlinearsystem}) in detail in Section 4 for $D_6$.
\paragraph{}
We also want to find the 'Levi-Civita connection', namely a regular torsion free and cotorsion free natural connection for a natural metric. We need for that end to find a suitable coframing or metric. As shown in \cite{riemanngeometry} a natural choice  in the group or quantum group case is to take any Ad-invariant nondegenerate bilinear form $\eta$ defined on $\Omega^*_0$, and indeed \cite{riemanngeometry} provides a ageneral 'braided-Killing  form' construction that can achieve this. Our 'cyclic' conjugacy class $\mathcal{C}$ described above is not, however, semi-simple and we instead have to determine all possible $\eta$.
\begin{theorem}
Let $\mathcal{C}$ be a cyclic conjugacy class with $n$ elements. Then up to normalisation, all nondegenerate $\mathrm{Ad}$-invariant bilinear forms on $\Omega^*_0$ are given by
\begin{equation}
\eta^{a,b}=\delta_{a,b}+\mu
\label{metric}
\end{equation}
for a constant $\mu \neq -1/n$. The associated metric in the Maurer-Cartan framing is
\begin{equation}
g=\sum_{a \in \mathcal{C}} e_a \underset{H}{\otimes } e_a+\mu  \theta \underset{H}{\otimes } \theta
\end{equation}
\end{theorem}
\textit{Proof}
\\
Here $g$ corresponds to an element $\eta \in \Omega_0 \otimes \Omega_0$ with coefficients $\eta^{a,b}$. We require it to be invertible (this is said  more abstractly  in \cite{riemanngeometry} to handle the quantum group case). The first condition is easily seen to be the requirement
\begin{equation}
\eta^{g^{-1}ag,b}=\eta^{a,gbg^{-1}}
\label{conjugaisonmetric}
\end{equation}
This and nondegeneracy is easy to see for the $\eta$ as stated.
Conversely, let us suppose that $\eta$ is
Ad-invariant and show that it is of form (\ref{metric}).
Since $\eta$ is Ad-invariant, it obeys (\ref{conjugaisonmetric}). We assume the existence of $t\in \mathcal{C}$ as in Definition 1, then $\mathrm{Ad}_t$ is a cyclic permutation of $\mathcal{C}-\{t\}$. From invariance (\ref{conjugaisonmetric}), it is obvious that 
$\eta^{t,a}=\eta^{t,\mathrm{Ad}_t(a)}=\eta^{t,\mathrm{Ad}_{t^2}(a)}=\eta^{t,\mathrm{Ad}_{t^{n-2}}(a)}$ for any $a \neq t$, and hence by cyclicity
\begin{equation}
\eta^{t,b}=\mu_1, \hspace{5mm}\forall b \neq t
\label{constantemetric}
\end{equation}
for some constant  $\mu_1$. But also from cyclicity we know that for any $a\in \mathcal{C}$ there is an element $c\in \mathcal{C}$ such that  $a=ctc^{-1}$. Hence 
from (\ref{constantemetric}), we have also
\begin{equation}
\eta^{a,b}=\eta^{ctc^{-1},b}=\eta^{t,c^{-1}bc}=\mu_1, \hspace{3mm} \forall a \neq b
\end{equation}
so all off-diagonals are $\mu_1$. Similarly, we have  $\eta^{a,a}=\eta^{ctc^{-1},ctc^{-1}}=\eta^{t,t}=\mu_2$ for all $a\in \mathcal{C}$ by Ad-invariance, for some constant $\mu_2$. Thus $\eta^{a,b}=(\mu_2-\mu_1)\delta_{a,b}+\mu_1$, which has, up to an overall scaling, the stated. The remaining condition  on the parameter $\mu$ comes from the fact that $\eta$ is invertible. Finally, given $\eta$ we define
\begin{equation}
e^{*a}=\sum_{b\in \mathcal{C}} e_b \eta^{ba}
\end{equation}
as explained in \cite{riemanngeometry} for the associated  coframing, which corresponds to the metric $g$ as stated.
\begin{flushright}
$\spadesuit$
\end{flushright}
\paragraph{}
One can then observe that this metric is symmetric in the sense 
\begin{equation}
\wedge g=0
\end{equation}
The theorem clarifies the observation in \cite{riemanngeometry} for $S_3$ where $\eta^{a,b}=\delta^{a,b}$ is derived as the braided Killing form (up to normalisation) but it is explained that one may add a multiple $\mu  \theta \underset{H}{\otimes } \theta$ to the metric (without changing the connection and Riemannian curvature). We are now ready to describe torsion free and cotorsion free connections in our cyclic case.
\begin{proposition}
In the setting of Proposition 1 and 2 and for the coframing given by $\eta$ as above, the torsion free and cotorsion   free connections obey the following relations:
\begin{eqnarray}
\nonumber R_x(\gamma)^{-1}=R_t(\gamma)^{-1}=R_y(\gamma)^{-1}\\
\nonumber R_x(\beta)^{-1}=R_t(\beta)^{-1}=R_y(\beta)^{-1}\\
R_x(\alpha)^{-1}=R_t(\alpha)^{-1}=R_y(\alpha)^{-1}\\
\label{relationavecR}
\end{eqnarray}
where $\alpha,\beta,\gamma$ are as in Proposition 2.
\end{proposition}
\textit{Proof:}\\
As in \cite{riemanngeometry}, when the coframing is given by the framing and an Ad-invariant  $\eta$, one may easily compute the form of the contorsion. One has 
\begin{equation}
D_A e^{*a}=\sum_{b \in \mathcal{C}} \eta^{ba} \mathrm{d}e_b+\sum_{b \in \mathcal{C},c\in \mathcal{C}} \eta^{ba} e_{cbc^{-1}} \wedge A_c -\sum_{b,c \in \mathcal{C}} \eta^{ba} e_b  \wedge A_c
\end{equation}
as a special case of the quantum groups computation in \cite{riemanngeometry}. Since we suppose the connections to  be torsion free, equation (\ref{sommedesAzero}) holds, then (cancelling $\eta^{ba}$), vanishing of contorsion in equation (\ref{DAetoile}) can be written equaivariently as
\begin{equation}
\mathrm{d}e_a+\sum_{b\in \mathcal{C}}e_{bab^{-1}}\wedge A_b=0
\label{equMaurerCartan}
\end{equation}
$\forall a \in \mathcal{C}$. If we write equation (\ref{equMaurerCartan}) for $a=t,x,y$ respectively, using Table 1, equations (\ref{equ2formD6}) and the definition of $\eta$, we obtain the following system of equations:
\begin{eqnarray}
\nonumber e_t\wedge A_t +e_y\wedge A_x+e_x\wedge A_y-e_y \wedge e_x-e_x\wedge e_y=0\\
\nonumber e_y \wedge A_t+e_x\wedge A_x+e_t\wedge A_y+e_x\wedge e_t+e_y\wedge e_x-e_y\wedge e_t=0\\
 e_x\wedge A_t+e_t\wedge A_x+e_y\wedge A_y+e_y\wedge e_t+e_x\wedge e_y-e_x\wedge e_t=0
 \label{systequ2formavecA}
\end{eqnarray}
We can get the third  equation of the system (\ref{systequ2formavecA}) from the two other. We then solve only the first two equations of this system. For that end, we set
\begin{equation}
A_a=e_bA_a^{'b}
\end{equation}
$\forall a \in \mathcal{C}$, with summation understood for $b\in \mathcal{C}$, and where we set
\begin{equation}
A^{'t}_t=1+\alpha '; A^{'x}_x=1+\beta '; A_y^{'y}=1+\gamma '
\end{equation}
as above. We then proceed in the same manner as we solved system (\ref{systequ2formavecA}), using this time the right module structure of $\Omega^1(H)$. We find that the solutions $(A_a)$ take the form
\begin{eqnarray}
\nonumber A^{t}=e_t (1+\alpha ')+e_x \gamma '+e_y \beta '\\
\nonumber A^{x}=e_t \gamma '+e_x (1+\beta ')+e_y \alpha '\\
 A^{y}=e_t \beta '+e_x \alpha '+e_y (1+\gamma ')
\label{definitiondeA'}
\end{eqnarray}
we then  wrtie these solutions using the left module structure on 1-form via the commutation relation in (\ref{relationcommutation}). And we compare the result to that of system (\ref{definitiondeA}) to obtain (\ref{relationavecR}) as stated.
\begin{flushright}
$\spadesuit$
\end{flushright}
\paragraph{}
At this level, we get many torsion free and cotorsion free connections. As one can remark, these equations for the connection do not depend on the corfficient $\mu$ of  $\theta \underset{H}{\otimes } \theta$, just as was the case  for $S_3$ in \cite{riemanngeometry}.
Modulo these modes, we see that there  is an essentially unique form of invariant metric on $G$ and we have given some conditions for associated moduli of torsion free and 
cotorsion free regular connections.

\section{Riemannian geometry of $D_6$}
In this section, we specialise to the group $D_6$ and present stronger results that depend on its structure and not only on the cyclic form of the conjugacy class. The group is defined by 
\begin{equation}
D_6=\{e,r,r^2,r^3,r^4,r^5,s,sr,sr^2,sr^3,sr^4,sr^5 \hspace{3mm}|\hspace{3mm} r^6=s^2=e\}
\end{equation}
where $e$ is  the group identity (this should be not confused with the Maurer-Cartan 1-form). We also have the relation:
\begin{equation}
sr^a=r^{-a}s
\end{equation}
In previous section, we considered $t,x,y$ such that 
\begin{equation}
t=sr \hspace{3mm} x=sr^3 \hspace{3mm}y=sr^5
\end{equation}
We choose the conjugacy class
\begin{equation}
\mathcal{C}=\{t,x,y\}
\end{equation}
which is 'cyclic'. Indeed, we have 
\begin{eqnarray}
\nonumber \mathrm{Ad}_t(x)=txt^{-1}=srsr^3sr=sr^5=y\\
\nonumber \mathrm{Ad}_t(y)=tyt^{-1}=srsr^5sr=sr^3=x\\
\end{eqnarray}
and
\begin{eqnarray}
\nonumber \mathrm{Ad}_x(t)=xtx^{-1}=sr^3srsr^3=sr^5=y\\
\nonumber \mathrm{Ad}_y(t)=yty^{-1}=sr^5srsr^5=sr^3=x\\
\mathrm{Ad}_t(t)=t
\end{eqnarray}
which show that $\mathcal{C}$ obeys conditions of Definition 1.
\paragraph{}
One may also check that the multiplication table of $\mathcal{C}$ correspond to table 2  as announced, so that we have at least one regular torsion free and cotorsion free connection on the bundle $H \otimes H$ where $H$ denotes from now $\C [D_6]$.

\begin{proposition}
For the cyclic conjugacy class on $D_6$, framing defined by the Maurer-Cartan form and coframing $e^*$ by Ad-invariant $\eta$, there exists a unique 'Levi-Civita' connection with component 1-forms
\begin{equation}
A_a=e_a-\frac{1}{3}\theta, \hspace{3mm}\forall a \in \mathcal{C}
\label{connection}
\end{equation}
\end{proposition}
\textit{Proof:}\\
The connection defined in (\ref{connection}) is easily seen to be solution of systems 
(\ref{definitiondeA}) and (\ref{definitiondeA'}). We are going to show that it is unique torsion free and cotorsion free connection which is solution of system (\ref{nonlinearsystem}). Using the properties of operators $(R_g)_{g \in D_6}$ and equations of system (\ref{relationavecR}) we find that (\ref{nonlinearsystem})  is equivalent to
\begin{eqnarray}
\nonumber -\gamma R_t(\beta)+(1+\beta)R_x(1+\alpha)-\beta R_t(\alpha)+\alpha R_x(\gamma)-(1+\alpha)R_t(1+\gamma)+\gamma R_x(\beta)=0\\
\nonumber -\gamma R_t(\beta)+\alpha R_y(\gamma)-\beta R_t(\alpha)+(1+\gamma)R_y(1+\beta)-(1+\alpha)R_t(1+\gamma)+\beta R_y(\alpha)=0\\
\nonumber -\gamma R_t(\gamma)+\alpha R_y(1+\alpha)-\beta R_t(1+\beta)+(1+\gamma)R_y(\gamma)-(1+\alpha)R_t(\alpha)+\beta R_y(\beta)=0\\
\nonumber -\gamma R_t(\gamma)+(1+\beta)R_x(\beta)-\beta R_t(1+\beta)+\alpha R_x(\alpha)-(1+\alpha)R_t(\alpha)+\gamma R_x(1+\gamma)=0\\
\nonumber -\beta R_t(\beta)+\alpha R_x(1+\alpha)-(1+\alpha)R_t(\alpha)+\gamma R_x(\gamma)-\gamma R_t (1+\gamma)+(1+\beta)R_x(\beta)=0\\
\nonumber -\beta R_t(\beta)+(1+\gamma)R_y(\gamma)-(1+\alpha)R_t(\alpha)+\beta R_y(1+\beta)-\gamma R_t(1+\gamma)+\alpha R_y(\alpha)=0\\
\nonumber -\beta R_t(\gamma)+(1+\gamma)R_y(1+\alpha)-(1+\alpha)R_t(1+\beta)+\beta R_y(\gamma) -\gamma R_t(\alpha)+\alpha R_y(\beta)=0\\
\nonumber -\beta R_t(\gamma)+\alpha R_x(\beta)-(1+\alpha)R_t(1+\beta)+\gamma R_x(\alpha)-\gamma R_t(\alpha)+(1+\beta)R_x(1+\gamma)=0
\end{eqnarray}

which have for unique solution $\alpha=\beta=\gamma=-\frac{1}{3}$. the expression of the corresponding connection (\ref{definitiondeA}) is then as stated. To end the proof of the Proposition 5, one checks easily that this connection is also solution of the other equations of system (\ref{definitiondeA'}).
\begin{flushright}
$\spadesuit$
\end{flushright}
We refer to this connection as \textit{the } 'Levi-Civita connection' for the invariant metric on the group $D_6$.
\begin{proposition}
The covariant derivative $\nabla:\Omega^1(H) \longrightarrow \Omega^1(H)\underset{H}{\otimes } \Omega^1(H)$ for the above 'Levi-Civita connection' on $D_6$, and its Riemann curvature $\mathcal{R}:\Omega^1(H)\longrightarrow \Omega^2 \underset{H}{\otimes } \Omega^1(H)$ are given by
\begin{eqnarray}
\nonumber \nabla (e_t)=-e_t  \underset{H}{\otimes } e_t-e_y  \underset{H}{\otimes }e_x-e_x  \underset{H}{\otimes }e_y+\frac{1}{3} \theta  \underset{H}{\otimes } \theta\\
\nonumber \nabla (e_x)=-e_y  \underset{H}{\otimes } e_t-e_x  \underset{H}{\otimes }e_x-e_t  \underset{H}{\otimes }e_y+\frac{1}{3} \theta  \underset{H}{\otimes } \theta\\
\nabla (e_y)=-e_x  \underset{H}{\otimes } e_t-e_t  \underset{H}{\otimes }e_x-e_y \underset{H}{\otimes }e_y+\frac{1}{3} \theta  \underset{H}{\otimes } \theta
\label{relationavecnabla}
\end{eqnarray}
and
\begin{eqnarray}
\nonumber \mathcal{R}(e_t)=\mathrm{d}e_t  \underset{H}{\otimes } e_t+\mathrm{d}e_y  \underset{H}{\otimes } e_x+\mathrm{d}e_x  \underset{H}{\otimes } e_y \\
\nonumber \mathcal{R}(e_x)=\mathrm{d}e_y  \underset{H}{\otimes } e_t+\mathrm{d}e_x  \underset{H}{\otimes } e_x+\mathrm{d}e_t  \underset{H}{\otimes } e_y \\
\mathcal{R}(e_y)=\mathrm{d}e_x  \underset{H}{\otimes } e_t+\mathrm{d}e_t  \underset{H}{\otimes } e_x+\mathrm{d}e_y  \underset{H}{\otimes } e_y
\label{relationavecR} 
\end{eqnarray}
\end{proposition}
\textit{Proof:}\\
The curvature 2-form $F$ is defined by equation (\ref{F}). In the present case, we have $bc \notin \mathcal{C}, \forall b,c \in \mathcal{C}$, so that $\sum_{b,c \in \mathcal{C},bc=a} A_b \wedge A_c=0,\forall a \in \mathcal{C}$. We have also  $\sum_{a \in \mathcal{C}}A_a=0$ and d$\theta=0$, hence $F_a=$d$A_a=$d$e_a$ for the form of the connection in (\ref{connection}). This is exactly the same argument as for $S_3$ in \cite{riemanngeometry}. Next, if we replace $\alpha$ in formula (\ref{R}) by $e_t,e_x,e_y$ respectively, and use Table 1, we obtain relation  (\ref{relationavecR}) for the curvature. Finally, we compute the value of the covariant derivative on the basis 1-form $\{e_a \}$ by using formula (\ref{nablaalpha}). Explicitly, we have 
\begin{eqnarray}
\nonumber \nabla (e_a)&=&-\sum_{b \in \mathcal{C}} A_b \underset{H}{\otimes } (e_{b^{-1}ab}-e_a)\\
\nonumber &=&-\sum_{b \in \mathcal{C}} (e_b-\frac{1}{3}\theta) \underset{H}{\otimes }(e_{b^{-1}ab}-e_a)\\
\nonumber &=&-\sum_{b \in \mathcal{C}} e_b \underset{H}{\otimes }(e_{b^{-1}ab}-e_a)+\frac{1}{3}\theta \underset{H}{\otimes }(e_{b^{-1}ab}-e_a)\\
\nonumber &&-\sum_{b \in \mathcal{C}} e_b \underset{H}{\otimes }e_{b^{-1}ab}+\sum_{b \in \mathcal{C}} e_b \underset{H}{\otimes } e_a+\frac{1}{3}\theta \underset{H}{\otimes }(e_{b}-e_a)\\
&&-\sum_{b \in \mathcal{C}} e_b \underset{H}{\otimes }e_{b^{-1}ab}+\frac{1}{3}\theta \underset{H}{\otimes }\theta
\end{eqnarray}
According to Table 1, this last equation gives relations (\ref{relationavecnabla}) as stated
\begin{flushright}
$\spadesuit$
\end{flushright}
From the Riemann curvature and the canonical lift $i$ we can compute the ricci   curvature of the Levi-Civita connection on $D_6$ and find that it vanishes. In fact we can prove a slightly stronger result  that it is the \textit{only} Ricci flat connection for this choice of framing.
\begin{theorem}
For the framing defined by the Maurer-Cartan 1-form, and for the canonical lift $i$, the above Levi-Civita connection on $D_6$  is the unique regular Ricci-flat connection.
\end{theorem}
\textit{Proof:}\\
In the present case, the canonical lift takes the form
\begin{equation}
i(e_a \wedge e_b)=e_a \underset{H}{\otimes} e_b-\frac{1}{2}\sum_{cd=ab,c \neq d} e_c \underset{H}{\otimes } e_d, \hspace{3mm}i(e_a \wedge e_a)=0
\end{equation}
We have to solve for vanishing of \cite{riemanngeometry}
\begin{equation}
Ricci=\sum_{a\in \mathcal{C}} <f^a,(i\otimes id)\mathcal{R}(e_a)>=\sum_{a,b \in \mathcal{C}} i(F_c)^{ab} e_b \underset{H}{\otimes} (e_{c^{-1}ac}-e_a)
\end{equation}
where $i(F_c)=i(F_c)^{ab}e_a \underset{H}{\otimes} e_b$, and the pairing is made between each $f^a$ and the first factor of the tensor product $(i \underset{H}{\otimes} id)\mathcal{R}(e_a)$ according to the formula $<f^a, me_b>=m\delta^a_b, \forall m\in H$. In our case this becomes
\begin{eqnarray}
\nonumber <f^t,i(F_x)\underset{H}{\otimes} (e_y-e_t)+i(F_y)\underset{H}{\otimes} (e_x-e_t)>\\
+\nonumber <f^x,i(F_t)\underset{H}{\otimes} (e_y-e_x)+i(F_y)\underset{H}{\otimes} (e_t-e_x)>\\
<f^y,i(F_t)\underset{H}{\otimes} (e_x-e_y)+i(F_x)\underset{H}{\otimes} (e_t-e_y)>
\label{Ricci}
\end{eqnarray}
We first compute $F_t,F_x,F_y$ :
\begin{eqnarray}
\nonumber F_t&=&\partial^t \alpha e_t \wedge e_t+\partial^x \alpha e_x \wedge e_t+\partial^y \alpha e_y\wedge e_t \\
\nonumber &&+\partial^t \gamma e_t \wedge e_x +\partial^x \gamma e_x \wedge e_x +\partial^y \gamma e_y \wedge e_x\\
\nonumber &&+\partial^t \beta e_t \wedge e_y+\partial^x \beta e_x \wedge e_y+ \partial^y \beta e_y \wedge e_y\\
\nonumber F_x&=&\partial^t \gamma e_t \wedge e_t+\partial^x \gamma e_x \wedge e_t+\partial^y \gamma e_y\wedge e_t \\
\nonumber &&+\partial^t \beta e_t \wedge e_x +\partial^x \beta e_x \wedge e_x +\partial^y \beta e_y \wedge e_x\\
\nonumber &&+\partial^t \alpha e_t \wedge e_y+\partial^x \alpha e_x \wedge e_y+ \partial^y \alpha e_y \wedge e_y\\
\nonumber F_y&=&\partial^t \beta e_t \wedge e_t+\partial^x \beta e_x \wedge e_t+\partial^y \beta e_y\wedge e_t \\
\nonumber &&+\partial^t \alpha e_t \wedge e_x +\partial^x \alpha e_x \wedge e_x +\partial^y \alpha e_y \wedge e_x\\
\nonumber &&+\partial^t \gamma e_t \wedge e_y+\partial^x \gamma e_x \wedge e_y+ \partial^y \gamma e_y \wedge e_y\\
\end{eqnarray}
and $i(F_t),i(F_x),i(F_y)$ for general free torsion connections in Proposition 2, then we rewrite equation (\ref{Ricci}) in terms of the basic elements $\{e_a \underset{H}{\otimes} e_b\}_{a,b \in \mathcal{C}}$ of the left H-module $\Omega^1(H)\underset{H}{\otimes} \Omega^1(H)$, the vanishing of each their 'first order derivatives' $\partial^a \alpha,\partial^a \beta,\partial^a \gamma; a\in \mathcal{C}$.
\paragraph{}
We find that it's enough to solve the following 4 equations coming from coefficients of $e_t \underset{H}{\otimes} e_t, e_x \underset{H}{\otimes} e_x, e_y \underset{H}{\otimes} e_y$ respectively:
\begin{eqnarray}
\nonumber -2\alpha +\beta +\gamma +\partial^t \beta +\partial^t \gamma-2\partial^x \beta +\partial^x \alpha +\partial^y \alpha -2\partial^y \gamma=0\\
\nonumber \alpha -2\beta +\gamma +\partial^x \alpha +\partial^x \beta +\partial^t \beta -2\partial^t \gamma +\partial^y \gamma -2\partial^y \alpha=0\\
\alpha +\beta -2\gamma +\partial^y \alpha +\partial^y \gamma +\partial^t \gamma -2\partial^t \beta +\partial^x \beta -2\partial^x \alpha =0
\label{equcoeffderivee}
\end{eqnarray}
Indeed, the unique solution of the mentioned system which obeys the condition  $\alpha +\beta +\gamma =-1$ as in Proposition 2 is $\alpha =\beta =\gamma=-\frac{1}{3}$. To end the  proof one just checks easily that this solution is also a solution of the 6 remainin equations (of the 9 ones mentioned above), coming from the coefficients of $e_a \underset{H}{\otimes} e_b, a \neq b$ in equation (\ref{Ricci})
\begin{flushright}
$\spadesuit$
\end{flushright}
One can also check that the Ricci tensor for the Levi-Civita connection with respect to the alternative 'lift'
\begin{equation}
i'(e_a \wedge e_b)=e_a \underset{H}{\otimes} e_b -e_{aba^{-1}} \underset{H}{\otimes} e_a
\end{equation}
also vanishes, i.e. the result does not depend strongly on the choice of lift. This is the same as found for $S_3$, where the two Ricci  tensors with respect  to $i$ and $i'$ respectively are the same up to a scale \cite{riemanngeometry}

\section{The Dirac operator for $D_6$}
Following the formalism of reference  \cite{riemanngeometry}, we write down in this section the 'gamma-matrices' and the Dirac operator  associated to the Maurer-Cartan framing $e$ and the  coframing $e^*$ for the invariant metric. We use the associated Levi-Civita connection constructed above.
\paragraph{}
For  the 'spinor' representation, we consider the standard 3-dimensional representation of $D_6$ defined  on a vector space  $W$ by
 \begin{eqnarray}
\nonumber
\rho (r) =
\left(\begin{array}{cc}
\omega &0\\
0&\omega^{-1}\\ 
   \end{array}
\right)&,& \hspace{5mm}
\rho (s) =
\left(\begin{array}{cc}
0 &1\\
1&0\\ 
   \end{array}
\right)
\\ \nonumber
\\
\rho (t) =
\left(\begin{array}{cc}
0&\omega^{-1}\\
\omega &0\\ 
   \end{array}
\right), \hspace{5mm}
\rho (x) &=&
\left(\begin{array}{cc}
0 &\omega^{-3}\\
\omega^3 &0\\ 
   \end{array}
\right)
\hspace{5mm}
\rho (x) =
\left(\begin{array}{cc}
0 &\omega^{-5}\\
\omega^5 &0\\ 
   \end{array}
\right)
\label{representation}
\end{eqnarray}
, $e$ is the unit matrix $I$ and $\omega=e^{i\frac{\pi}{3}}$. The Casimir element $C$ associated to the operator $\eta$ is given in \cite{riemanngeometry} by
\begin{equation}
C=\eta_{ab}^{-1} f^a f^b=\eta_{ab}^{-1}  (a-e)(b-e)
\end{equation}
with summation understood $b,a \in \mathcal{C}$. One checks that it corresponds in the general case of the class $\mathcal{C}=\{t,x,y\}$ as  in Proposition 1, to the explicit form
\begin{equation}
C=\frac{1}{1+4\mu}  [(t-e)^2+(x-e)^2+(y-e)^2]
\end{equation}
then 
\begin{equation}
\rho_W(C)=\frac{6}{1+4\mu}I
\end{equation}
Next, we choose our gamma-matrix $\gamma$ to be the tautological gamma-matrix' \cite{riemanngeometry} associated to $\rho_W$ and $\eta$ defined by
\begin{equation}
\gamma_a=\eta^{-1}_{ab} \rho_W(f^b)=\sum_{b \in \mathcal{C}} \eta^{-1}_{ab} \rho_W (b-e), \hspace{3mm} \forall a \in \mathcal{C}
\label{gammaeta}
\end{equation}
In our case we find
\begin{equation}
\gamma_a=\rho_W(a-e)+\frac{4\mu}{1+4\mu}
\label{defgamma}
\end{equation}
and that these matrices obey the relations 
\begin{equation}
\sum_{a \in \mathcal{C}}  \gamma_a=-\frac{3}{1+4\mu}
\label{sommegammaa}
\end{equation}
following directly from (\ref{defgamma}).
\paragraph{}
Equation (\ref{defgamma})and (\ref{sommegammaa}) hold in general case considered in Proposition 1, providing that the multiplication's table is that of Table 2.  The explicit matrix representation of these gamma-matrices above for $D_6$ are
  \begin{eqnarray}
\nonumber
\gamma_i =
\left(\begin{array}{cc}
-\frac{1}{1+4\mu} &\omega^{-1-2i}\\
\omega^{1+2i}&-\frac{1}{1+4\mu}\\ 
   \end{array}
\right)
\label{repgamma}
\end{eqnarray}
where $i=0,1,2$ correspond to $t,x,y$ respectively.
\begin{proposition}
The Dirac operator (\ref{defdiracoperator}) on $D_6$ for the gamma-matrices and the Levi-Civita connection on $D_6$ constructed above is given by
\begin{equation}
\cancel{D}=\partial^a \gamma_a-3
\label{resultatoperateurdirac}
\end{equation}
(sum over $a \in \mathcal{C}$). For $\mu=0$, we have explicitly
 \begin{eqnarray}
\nonumber
\cancel{D}=
\left(\begin{array}{cc}
-R_t-R_x-R_y &\omega^{-1}R_t+\omega^{-3}R_x+\omega^{-5}R_y\\
\omega R_t+\omega^3R_x+\omega^5 R_y&-R_t-R_x-R_y\\ 
   \end{array}
\right)
\label{repgamma}
\end{eqnarray}
This has 8 zero modes, 8 modes with eigenvalue +3, 8 modes with eigenvalues -3
\end{proposition}
\textit{Proof:}\\
The formula giving the Dirac operator in terms of the gamma-matrices and the representation $\rho_W$ is given by equation (\ref{defdiracoperator}). We first  observe that for  the representation $\rho_W$ above, the following two equations hold:
$\sum_{a \in \mathcal{C}} \rho_W(a)=0$ and $\sum_{a \in \mathcal{C}} \rho_W(a^2)=0$. Using the $A^b_a$ defined by (\ref{connection}), and the fact that every  element  of $\mathcal{C}$ is of order 3, we obtain 
\begin{eqnarray}
\nonumber \cancel{D}&=&\partial^a \gamma_a-\sum_{a,b \in \mathcal{C}}(\delta_a^b-\frac{1}{3})[\rho_W(a-e)+\frac{4\mu-1}{1+4\mu}] \rho_W(b-e)\\
\nonumber &=&\partial^a \gamma_a-\sum_{a \in \mathcal{C}}[\rho_W(a-e)+\frac{4\mu-1}{1+4\mu}]\rho(a-e)+\frac{1}{3}\sum_{a \in \mathcal{C}}[\rho_W(a-e)+\frac{4\mu-1}{1+4\mu}](-3I)\\
\nonumber &=&\partial^a-\sum_{a\in \mathcal{C}}[\rho_W(a^2-2a+e)-\frac{1\mu-1}{1+4\mu}\sum_{a\in \mathcal{C}}\rho(a-e)-\sum_{a\in \mathcal{C}}\\
\nonumber &=&\partial^a \gamma_a-6I+\frac{3(4\mu-1)}{1+4\mu}+\frac{6}{1+4\mu}\\
\nonumber &=&\partial^a \gamma_a-3
\end{eqnarray}
We then replace in equation (\ref{resultatoperateurdirac}) the representation of the gamma-matrices from (\ref{repgamma}) to obtain the matrix representation of $\cancel{D}$ as stated.
\paragraph{}
To compute its eigenvalues we need $R_a$ explicitly as $12\times 12$ matrices. In the basis spanned by delta-functions at $\{e,r,r^2,r^3,r^4,r^5,s,sr,sr^2,sr^3,sr^4,sr^5\}$, the right translation operators take the form
 \begin{eqnarray}
R_j =
\left(\begin{array}{ccc}
0_3&J&0_4\\
^tJ&0_6&^tJ\\
0_4&J&0_3\\ 
   \end{array}
\right)
\label{matriceR}
\end{eqnarray}
where $0_n$ is the $n\times n$-square zero matrix and $j=t,x,y$ which correspond to $J=T,X,Y$ the matrices given by:
 \begin{eqnarray}
T=
\left(\begin{array}{ccccc}
1&0&0&0&0\\
0&0&0&0&1\\
0&0&1&0&0\\ 
   \end{array}
\right)
X=
\left(\begin{array}{ccccc}
0&0&1&0&0\\
1&0&0&0&0\\
0&0&0&0&1\\ 
   \end{array}
\right)
Y=
\left(\begin{array}{ccccc}
0&0&0&0&1\\
0&0&1&0&0\\
1&0&0&0&0\\ 
   \end{array}
\right)
\label{fonctiondematriceR}
\end{eqnarray}
We then obtain the eigenvalues as stated
\begin{flushright}
$\spadesuit$
\end{flushright}
\paragraph{}
The eigenvalues here do in fact depend on $\mu$ and the case $\mu=0$ seems to be the more natural since it corresponds to the simplest metric $\delta_{a,b}$. The $-3$ in (\ref{resultatoperateurdirac}) corresponds to the constant curvature of $D_6$ as for $S_3$ in \cite{riemanngeometry}. As for $S_3$, this offset ensures a symmetrical distribution of eigenvalues about zero.
\paragraph{}
We will now construct the eigenstates of $\cancel{D}$. Before doing that we look at the spin $0$ or scalar wave equation defined by the corresponding wave operator
\begin{equation}
\square =-\eta_{ab}^{-1} \partial^a \partial^b=-\sum_a \partial^a  \partial^a=\sum_a(2R_a-R_{a^2}-\mathrm{id})
\end{equation}
We do not exactly expect a Lichnerowicz formula relating this to the square of $\cancel{D}$, but we find that it is the square of a first-order operator with eigenvalues contained in those of $\cancel{D}$. It is easy to solve the wave equation directly.
\begin{proposition}
\begin{equation}
\square =2D_0-6\mathrm{id}, \hspace{5mm} D_0=\sum_a R_a
\end{equation}
There are 2 zero modes given by a two dimensional representation $\tilde{\rho}_{k2}$ with $1\le k\le 2$. There are 2 modes with eigenvalues $-12$ given by a two dimensional representation $\tilde{\rho}_{k1}$ with $1\le k\le 2$. There are $8$ modes given by  $2$ other representation of $D_6$: $\rho_{kl}$ and its complex conjugate $\overline{\rho}_kl$ where $1\le k,l\le 2$.
\end{proposition}
\textit{Proof:}
The square form of $\square$ follows from the multiplication Table $2$. From there one finds that $\sum_a R_{a^2}=\sum_a R_{t^2}=3\mathrm{id}$, after which the result follows.
To solve the wave equation, we introduce the first two dimensional representation
 \begin{eqnarray}
\tilde{\rho} (s) =
\left(\begin{array}{ccc}
-1&0\\
0&1\\ 
   \end{array}
\right), \hspace{5mm}\tilde{\rho} (r)=\mathrm{id}
\label{premiererepresentation}
\end{eqnarray}
Then we have $\tilde{\rho} (s)=\tilde{\rho} (t)=\tilde{\rho} (x)=\tilde{\rho} (y)$.
Then $\forall m \in \mathcal{C}$, 
\begin{eqnarray}
\nonumber \square \tilde{\rho}_{k1} (m)&=&\sum_a[2R_a \tilde{\rho}_{k1}(m)-R_{a^2}\tilde{\rho}_{k1}(m)-\tilde{\rho}_{k1}(m)] \\
\nonumber &=&\sum_a \sum_i[2 \tilde{\rho}_{ki}(m)\tilde{\rho}_{i1}(a)-\tilde{\rho}_{ki}(m)  \tilde{\rho}_{i1}(a^2)]-3\tilde{\rho}_{k1}(m) \\
\nonumber &=&\sum_a \sum_i[2 \tilde{\rho}_{ki}(m)\tilde{\rho}_{i1}(a)-\tilde{\rho}_{ki}(m)  \tilde{\rho}_{i1}(a^2)]-3\tilde{\rho}_{k1}(m) \\
\nonumber &=& -\sum_i[9\tilde{\rho}_{ki}(m)\mathrm{id}^{i1}]-3\tilde{\rho}_{k1}(m) \\
 \square \tilde{\rho}_{k1} (m)&=&-12\tilde{\rho}_{k1}(m)
\end{eqnarray}
since $\sum_a \tilde{\rho}_{k1}(a^2)=-\sum_a \tilde{\rho}_{k1}(a)=3\mathrm{id}$. Similary, for the matrix element $\{ \tilde{\rho}_{k2}\}$, we have:
\begin{eqnarray}
\nonumber \square \tilde{\rho}_{k2} (m)&=&\sum_a[2R_a \tilde{\rho}_{k2}(m)-R_{a^2}\tilde{\rho}_{k2}(m)-\tilde{\rho}_{k2}(m)] \\
\nonumber &=&\sum_a \sum_i[2 \tilde{\rho}_{ki}(m)\tilde{\rho}_{i2}(a)-\tilde{\rho}_{ki}(m)  \tilde{\rho}_{i2}(a^2)]-3\tilde{\rho}_{k2}(m) \\
\nonumber &=&\sum_a \sum_i[2 \tilde{\rho}_{ki}(m)\tilde{\rho}_{i2}(a)-\tilde{\rho}_{ki}(m)  \tilde{\rho}_{i2}(a^2)]-3\tilde{\rho}_{k2}(m) \\
\nonumber &=& \sum_i[3\tilde{\rho}_{ki}(m)\mathrm{id}^{i2}]-3\tilde{\rho}_{k2}(m) \\
 \square \tilde{\rho}_{k2} (m)&=&0
\end{eqnarray}
since $\sum_a \tilde{\rho}_{k1}(a^2)=\sum_a \tilde{\rho}_{k1}(a)=3\mathrm{id}$
Now we calcul the eigenvalues of $\rho_{kl}(m)$
\begin{eqnarray}
\nonumber \square \rho_{kl} (m)&=&\sum_a[2R_a \rho_{kl}(m)-R_{a^2}\rho_{kl}(m)-\rho_{kl}(m)] \\
\nonumber &=&\sum_a \sum_i[2 \rho_{ki}(m)\rho_{il}(a)-\rho_{ki}(m)  \rho_{il}(a^2)]-3\rho_{kl}(m) \\
\nonumber &=&\sum_a \sum_i[2 \rho_{ki}(m) \rho_{il}(a)-\rho_{ki}(m) \rho_{il}(a^2)]-3\rho_{kl}(m) \\
\nonumber &=& -\sum_i[3\rho_{ki}(m)\mathrm{id}^{il}]-3\rho_{kl}(m) \\
 \square \rho_{kl} (m)&=&-6\rho_{kl}(m)
\end{eqnarray}
and the same for $ \overline{\rho}_{kl} (m)$ instead of $ \rho_{kl} (m)$.
\begin{flushright}
$\spadesuit$
\end{flushright}
Moreover, every function $\phi$ on $D_6$ has a unique decomposition of the form
\begin{equation}
 \phi=\sum_{k}p_k \tilde{\rho}_{k1}+\sum_{k}q_k \tilde{\rho}_{k2}+\sum_{kl}p_{kl} \tilde{\rho}_{kl}+\sum_{kl}q_{kl} \tilde{\rho}_{kl}
\end{equation}
for  some numbers $p_k,q_k,p_{kl},q_{kl}$ which are components of $\phi$ in the nonabelian Fourier transform. The decomposition above corresponds precisely to the Peter-Weyl decomposition, just as noted for $S_3$ in \cite{Strois}.
\paragraph{}
We now use the preceding results to completely solve the Dirac equation.
We set 
\begin{equation}
D_1=\omega R_t+\omega^3 R_x+\omega^5 R_y, \hspace{5mm}D_2=\omega^{-1}R_t+\omega^{-3}R_x+\omega^{-5}R_y
\end{equation}
so that
 \begin{eqnarray}
\cancel{D} =
\left(\begin{array}{cc}
-D_0&D_2\\
D_1&-D_0\\ 
   \end{array}
\right)
\label{DbarreenfonctiondeDoetDun}
\end{eqnarray}
Let us note first of all that
\begin{equation}
D_1^2=D_2^2=0 \hspace{5mm}D_0D_2=D_0D_1=D_1D_0=D_2D_0=0
\end{equation}
from which we see by inspection that the following are 8 linearly-independant zero modes of $\cancel{D}$
 \begin{eqnarray}
\left(\begin{array}{c}
D_1 \rho_{k1}\\
0\\ 
   \end{array}
\right)
\left(\begin{array}{c}
D_1 \rho_{k2}\\
0\\ 
   \end{array}
\right)
\left(\begin{array}{c}
0\\
D_2 \rho_{k1}\\ 
   \end{array}
\right)
\left(\begin{array}{c}
0\\
D_2 \rho_{k2}\\ 
   \end{array}
\right)
\label{zeromodes}
\end{eqnarray}
for $1\le k\le 2$. Similarly it is immediate by inspection that 
 \begin{eqnarray}
\left(\begin{array}{c}
 \tilde{\rho}_{k1}\\
0\\ 
   \end{array}
\right)
\left(\begin{array}{c}
0\\
\tilde{\rho}_{k1}\\ 
   \end{array}
\right)
\label{troismodes}
\end{eqnarray}
are 4 modes with eigenvalue 3. This is because $D_0 \tilde{\rho}_{k1}=3\tilde{\rho}_{k1}$ (as in Proposition 8 above) while $D_1 \tilde{\rho}_{k1}=D_2 \tilde{\rho}_{k1}=0$.
\paragraph{}
We also have that 
 \begin{eqnarray}
\left(\begin{array}{c}
 \tilde{\rho}_{k2}\\
0\\ 
   \end{array}
\right)
\left(\begin{array}{c}
0\\
\tilde{\rho}_{k2}\\ 
   \end{array}
\right)
\label{troismodes}
\end{eqnarray}
are 4 modes with eigenvalue -3. This is because $D_0 \tilde{\rho}_{k2}=-3\tilde{\rho}_{k2}$ (as in Proposition 8 above) while $D_1 \tilde{\rho}_{k2}=D_2 \tilde{\rho}_{k2}=0$.
\paragraph{}
It remains to construct 4 modes xith eigenvalues +3 and -3. Before doing this let us make two observations about the modes already evident. First of all, let $\hat{\tilde{\rho}}_{k1}$ denote the operator of multiplication by $\tilde{\rho}_{k1}$. Then $R_a \hat{\tilde{\rho}}_{k1}=-\hat{\tilde{\rho}}_{k1}R_a$ since 
 \begin{eqnarray}
 \tilde{\rho} (a)=
\left(\begin{array}{cc}
-1&0\\
0&1\\ 
   \end{array}
\right)
\label{representationtriviale}
\end{eqnarray}
for $a=t,x,y$. Hence
\begin{equation}
\cancel{D} \hat{\tilde{\rho}}_{k1}=-\hat{\tilde{\rho}}_{k1} \cancel{D}
\end{equation}
Thus, multiplication of a spinor mode by function $\tilde{\rho}_{k1}$ multiplies its eigenvalue by $-1$. This generates the -3 modes above from the $n=0$ case.
\\
Secondly, from the multiplication Table 2 we see that 
\begin{eqnarray}
\nonumber R_t D_1=\omega^2 D_2 R_t\\
R_t D_2=\omega^{-2} D_1 R_t
\end{eqnarray}
Note that
\begin{eqnarray}
\nonumber R_t \rho_{k2}=\omega^{-1} \rho_{k1}\\
R_t \rho_{k1}=\omega \rho_{k2}
\end{eqnarray}
from the explicit form of $\rho (t)$.
\\
We now observe that if we make an ansatz of the form
\begin{equation}
\psi =\left(\begin{array}{c}
\phi \\
\omega R_t \phi  \\
   \end{array}
\right)
=\left(\begin{array}{c}
\mathrm{id} \\
\omega R_t  \\
   \end{array}
\right)\phi
\end{equation}
for function $\phi$ then 
\begin{equation}
\cancel{D}\psi=\left(\begin{array}{c}
\mathrm{id} \\
\omega R_t  \\
   \end{array}
\right)(-D_0+\omega D_2 R_t)\phi
\end{equation}
so eigenspinors are induced by eigenfunctions of the operator
\begin{equation}
-D_0+\omega D_2 R_t=-D_0+R_e +\omega^4 R_{xt} +\omega^2 R_{yt}
\end{equation}
All the $\rho_{kl}$ are zero modes of $D_0$ (as in Proposition 8), while among them precisely $\rho_{k2}$  is an eigenmode of $R_e +\omega^4 R_{xt} +\omega^2 R_{yt}$, with eigenvalue 3 (this follows $\frac{1}{3}(\rho (e)+\omega^4 \rho (xt)+\omega^2 \rho  (yt))$ being a projection matrix of rank 1). Hence $\phi =\rho_{k2}$ in the ansatz yields three spinor modes
\begin{equation}
\left(\begin{array}{c}
\rho_{k2} \\
\rho_{k1}  \\
   \end{array}
\right), \hspace{5mm}1\le k \le 2
\end{equation} 
 with eigenvalue +3 of $\cancel{D}$. Applying  $\hat{\tilde{\rho}}_{k1}$ generates two with eigenvalue -3.

 \paragraph{}
 Secondly, from the multiplication Table 2 we see that 
\begin{eqnarray}
\nonumber R_x D_1= D_2 R_x\\
R_x D_2= D_1 R_x
\end{eqnarray}
Note that
\begin{eqnarray}
\nonumber R_x \rho_{k1}=- \rho_{k2}\\
R_x \rho_{k2}=-\rho_{k1}
\end{eqnarray}
from the explicit form of $\rho (t)$.
\\
We now observe that if we make an ansatz of the form
\begin{equation}
\psi =\left(\begin{array}{c}
\phi \\
R_x \phi \\
   \end{array}
\right)
=\left(\begin{array}{c}
\mathrm{id} \\
R_x  \\
   \end{array}
\right)\phi
\end{equation}
for function $\phi$ then 
\begin{equation}
\cancel{D}\psi=\left(\begin{array}{c}
\mathrm{id} \\
\omega R_x  \\
   \end{array}
\right)(-D_0+\omega D_2 R_x)\phi
\end{equation}
so eigenspinors are induced by eigenfunctions of the operator
\begin{equation}
-D_0+\omega D_2 R_t=-D_0-R_e +\omega^{-1} R_{yt} +\omega^{-5} R_{xt}
\end{equation}
All the $\rho_{k2}$ are zero modes of $D_0$ (as in Proposition 8), while among them precisely $\rho_{k2}$  is an eigenmode of $-R_e +\omega^{-1} R_{yt} +\omega^{-5} R_{xt}$, with eigenvalue -3 (this follows $\frac{1}{3}(\rho (e)-\omega^{-1} \rho (yt)-\omega^{-5} \rho  (xt))$ being a projection matrix of rank 1). Hence $\phi =\rho_{k2}$ in the ansatz yields three spinor modes
\begin{equation}
\left(\begin{array}{c}
\rho_{k2} \\
-\rho_{k1}  \\
   \end{array}
\right), \hspace{5mm}1\le k \le 2
\end{equation} 
 with eigenvalue -3 of $\cancel{D}$. Applying  $\hat{\tilde{\rho}}_{k1}$ generates two with eigenvalue +3.
\\
This completes our diagonalisation of $\cancel{D}$. Finally, we note that there necessarily exists an operator $\gamma$ with $\gamma^2=\mathrm{id}$ and $\{\gamma,\cancel{D}\}=0$, but it is not unique. Thus, diagonalising $\cancel{D}$,  we can group the eigenbasis into pairs of 2-blocks of zero modes according to the two groups in (\ref{zeromodes}), interchanged by $\gamma$, and similarly we define $\gamma$ to interchange the two 2-blocks with eigenvalues $\pm 3$. This defines at least one choice of $\gamma$, suggested by our explicit diagonalisation.

\section{Spectral action}
We now compute the spectral action from \cite{spectralaction}
\begin{eqnarray}
\nonumber \mathrm{Tr}[f(\frac{\cancel{D}^2}{\Lambda^2})]&=&\sum_{k=-1}^1 8 f(\frac{9k^2}{\Lambda^2})\\
\nonumber  8 \sum_{k=-1}^1 f(\frac{9k^2}{\Lambda^2})&=& 8 \sum_{k=-1}^1 \hat{f}(\frac{9k^2}{\Lambda^2})\\
\nonumber&=&8 \sum_{k=-1}^1 \int_{\R} f(u)e^{-2i\pi u \frac{9k^2 }{\Lambda^2}}du\\
\nonumber&=&8\int_{\R} f(u)(2e^{-2i\pi u \frac{9 }{\Lambda^2}}+1)du\\
\nonumber&=&16\int_{\R} f(u)e^{-2i\pi u \frac{9 }{\Lambda^2}}du+8\int_{\R} f(u)du\\
\mathrm{Tr}[f(\frac{\cancel{D}^2}{\Lambda^2})] &\simeq & 24\int_{\R} f(u)du-\frac{288 i\pi}{\Lambda^2}\int_{\R} uf(u)du-\frac{5384\pi^2}{\Lambda^4}\int_{\R} u^2 f(u)du
\end{eqnarray}
where $\hat{f}$ is the Fourier transform of $f$

\newpage

\section{Discussion}
We have chosen the Dihedral group $D_6$ because its conjugacy classis cyclic. But you ca remark that the conjugacy class $\mathcal{C}=\{sr,sr^3,sr^5\}$ is isomorph to the conjugacy class of $S_3$ taken in \cite{riemanngeometry} $\mathcal{C}=\{u,v,uvu\}$ i.e. it has the same multiplication table (Table 2).
\paragraph{}
However, the representations and so the Dirac operators  of $S_3$ and $D_6$ are differents. In \cite{riemanngeometry}, is eigenvalues are $\pm 1$ whereas for $D_6$ its eigenvalues are $\pm 3$. Furthermore, \cite{riemanngeometry} has not calculate the eigenmodes of the Dirac operators. But we have done  it in this paper.
\paragraph{}
The last section try to calculate the spectral action of the Dirac operator we have calculated. We used in the second line of the calculation the Poisson summation formula which is true for summation over $\Z$ but I supposed it true for summation over $-1,0,1$.

\end{document}